\documentclass[conference]{IEEEtran}

\title{Smooth R\'enyi Entropy of Ergodic Quantum Information Sources}

\author{
\authorblockN{Berry Schoenmakers \ \ \ \ \ \ \ Jilles Tjoelker}
\authorblockA{Dept.\ of Mathematics and Computer Science \\
Technical University Eindhoven\\ The Netherlands\\
{\footnotesize\tt berry@win.tue.nl} \ \ \ {\footnotesize\tt j.tjoelker@student.tue.nl}}
\and
\authorblockN{Pim Tuyls}
\authorblockA{Information Security Systems\\
Philips Research Eindhoven\\ The Netherlands\\
{\footnotesize\tt pim.tuyls@philips.com}}
\and
\authorblockN{Evgeny Verbitskiy}
\authorblockA{Digital Signal Processing\\
Philips Research Eindhoven\\ The Netherlands\\
{\footnotesize\tt evgeny.verbitskiy@philips.com}}}

\usepackage[english]{babel}
\usepackage{amsfonts}
\usepackage{amsmath}
\usepackage{amssymb}
\usepackage{bbm}

\newtheorem{definition}{Definition}
\newtheorem{theorem}{Theorem}
\newtheorem{lemma}{Lemma}

\newcommand{\card}[1]{|#1|} 
\newcommand{\bit}{\{0,1\}}

\newcommand{\tr}{\mathop{\mathrm{tr}}\nolimits}
\newcommand{\rank}{\mathop{\mathrm{rank}}\nolimits}

\newcommand{\pd}[1]{\mathbb{#1}}
\newcommand{\PP}{\pd{P}}
\newcommand{\QQ}{\pd{Q}}

\newcommand{\mN}{\mathbb{N}}

\newcommand{\SmoothRenyiEntropy}[3]{#1_{#2}^{#3}}
\newcommand{\SmoothRenyiEntropyRate}[3]{#1_{#2}^{#3}}
\newcommand{\Hrs}[2]{\SmoothRenyiEntropy{H}{#1}{#2}}
\newcommand{\Hr}[1]{\SmoothRenyiEntropy{H}{#1}{}}
\renewcommand{\H}{\SmoothRenyiEntropy{H}{}{}}
\newcommand{\hrs}[2]{\SmoothRenyiEntropyRate{h}{#1}{#2}}

\newcommand{\h}{\SmoothRenyiEntropyRate{h}{}{}}
\newcommand{\Srs}[2]{\SmoothRenyiEntropy{S}{#1}{#2}}
\newcommand{\Sr}[1]{\SmoothRenyiEntropy{S}{#1}{}}
\renewcommand{\S}{\SmoothRenyiEntropy{S}{}{}}
\newcommand{\srs}[2]{\SmoothRenyiEntropyRate{s}{#1}{#2}}

\newcommand{\s}{\SmoothRenyiEntropyRate{s}{}{}}

\renewcommand{\b}{e} 
\renewcommand{\exp}[1]{\b^{#1}}
\newcommand{\typ}{\mathcal{T}^{(n)}_\epsilon}
\newcommand{\ball}{\mathcal{B}^\epsilon}
\newcommand{\n}[1]{#1^{(n)}}

\begin{document}

\maketitle

\begin{abstract}
We investigate the recently introduced notion of smooth R\'enyi entropy for the case of
{\em ergodic} information sources, thereby generalizing previous work
which concentrated mainly on i.i.d.\ information sources. We will actually
consider ergodic {\em quantum} information sources, of which ergodic classical
information sources are a special case.
We prove that the average smooth R\'enyi entropy rate will approach the
entropy rate of a stationary, ergodic source, which is equal to
the Shannon entropy rate for a classical source and the von Neumann entropy rate for a
quantum source.
\end{abstract}

\section{Introduction}
The elegant notion of smooth R\'enyi entropy was introduced recently by
Renner and Wolf in \cite{RW04} for classical information sources, and the
natural extension to quantum information sources
was defined by Renner and K\"onig in \cite{RK05}. In these two papers and further
work by Renner and Wolf \cite{RW05,Ren05}, many properties of smooth
R\'enyi entropy---and smooth min-entropy and smooth max-entropy in particular---have been
studied in detail.

A central property of smooth R\'enyi entropy proved in these works is that for
memoryless (i.i.d.) information sources, the average smooth R\'enyi entropy rate will approach the
entropy rate of the source, which is equal to the Shannon entropy for a classical source
and the von Neumann entropy for a quantum source. Whereas, in general, the average
(conventional) R\'enyi entropy rate
of a memoryless source does not converge to the source's entropy rate.

In this paper we extend the study of smooth R\'enyi entropy to the more general class
of stationary, ergodic sources rather than memoryless sources. We will prove that
for both the classical and the quantum case that the average smooth R\'enyi entropy rate
will approach the Shannon and the von Neumann entropy rate, respectively. We will do
so by first treating the classical case and then reducing the quantum case to the
classical one without losing generality.

In general, smooth R\'enyi entropy of order $\alpha>1$, and $\alpha=\infty$ (min-entropy) in particular,
is of cryptographic relevance (e.g., for randomness-extraction), and smooth R\'enyi entropy of
order $\alpha<1$, and $\alpha=0$ (max-entropy) in particular, are relevant to data compression (minimum
encoding length). In these contexts, the importance of {\em smooth} R\'enyi entropy is that its rate
is basically equal to the Shannon/von Neumann entropy rate for an i.i.d.\ source (and for ergodic sources
as well, as we show in this paper). This is not the case for conventional R\'enyi entropy.
More generally, as shown in the papers by Renner {\em et al.}\ mentioned above, smooth R\'enyi
entropy behaves much as Shannon/von Neumann entropy does.

In this paper we focus on the unconditional case, whereas much of the abovementioned work by Renner {\em et al.}\
treats the more general conditional case.
We leave the extension to the conditional case for future work. However, we do consider
two notions of $\epsilon$-closeness, one based on trace distance (also known as variational
or statistical distance) and one based on non-normalized density matrices (or probability distributions),
where the latter is more suitable to handle the conditional case.\footnote{The trace distance was originally used
in \cite{RW04,RK05}. The use of non-normalized probability distributions was also shown in the full version
of \cite{RW04} and used in \cite{RW05}. In this paper, we extend this to the use of non-normalized density matrices
in the quantum case.}
Thus, we believe that our results can be extended to the conditional case as well.

We also note that Renner \cite{Ren05} presents a different kind of generalization of i.i.d.\ quantum
sources, namely by analyzing the smooth min-entropy of symmetric (permutation-invariant) quantum states. Or, more
precisely, states in a symmetric subspace of ${\cal H}^{\otimes n}$ are considered, for $n\in\mN$.
See \cite[Chapter~4]{Ren05} for details, which also covers the conditional case.


\section{Preliminaries}\label{sec:prelim}
Throughout this paper we use $\PP$ and $\QQ$ to denote
probability distributions with over the same finite or countably infinite range $\cal Z$.
Similarly, we use $\rho$ and $\sigma$ to denote density matrices
on the same Hilbert space of a finite or countably infinite dimension. These probability
distributions and density matrices are not necessarily normalized (e.g., $\sum_z \PP(z)<1$
if $\PP$ is non-normalized and $\tr(\rho)<1$ if $\rho$ is non-normalized).

For ease of comparison we state all the preliminaries explicitly for the classical case
as well as for the quantum case.

\begin{definition}[Classical R\'enyi entropy] The R\'enyi entropy of order
$\alpha\in[0,\infty]$ of probability distribution $\PP$ is
\[
\Hr{\alpha}(\PP) = \frac{1}{1-\alpha} \log \sum_{z\in\mathcal{Z}} \PP(z)^\alpha,
\]
for $0<\alpha<\infty$, $\alpha\neq1$, and
$\Hr{\alpha}(\PP) =\lim_{\beta\rightarrow\alpha} \Hr{\beta}(\PP)$ otherwise.
\end{definition}
Hence, $\Hr{0}(\PP) = \log\card{\{z\in\mathcal{Z}: \PP(z)>0\}}$, $\Hr{1}(\PP)=\H(\PP)$ (Shannon entropy)
and $\Hr{\infty}(\PP) = - \log \max_{z\in\mathcal{Z}}\PP(z)$.

For a random variable $Z$ we use $\Hr{\alpha}(Z)$ as a shorthand for
$\Hr{\alpha}(\PP_Z)$, where $\PP_Z$ is the probability distribution of $Z$.

Smooth R\'enyi entropy was introduced in \cite{RW04} for the classical case.
For $\epsilon\geq0$, let $\ball(\PP)$ denote either the set of probability
distributions which are $\epsilon$-close to $\PP$, $\ball(\PP)=\{\QQ: \delta(\PP,\QQ)\leq\epsilon\}$,
or the set of non-normalized probability distributions which are $\epsilon$-close
to $\PP$, $\ball(\PP)=\{\QQ: \sum_{z\in\mathcal{Z}}\QQ(z)\geq1-\epsilon,
\forall_{z\in\mathcal{Z}} 0\leq \QQ(z) \leq \PP(z)\}$. The first notion of $\epsilon$-closeness,
based on the statistical distance $\delta(\PP,\QQ)=\frac{1}{2}
\sum_{z\in\mathcal{Z}} |\PP(z)-\QQ(z)|$, was used in \cite{RW04}. The second notion
was mentioned in the full version of \cite{RW04}, and used in \cite{RW05}.
\begin{definition}[Classical smooth R\'enyi entropy, \cite{RW04}]\label{def:classicsmooth}
The $\epsilon$-smooth R\'enyi entropy of order $\alpha\in[0,1)\cup(1,\infty]$ of a
probability distribution $\PP$ is
\[
\Hrs{\alpha}{\epsilon}(\PP) = \left\{\begin{array}{ll}
    \inf_{\QQ\in\ball(\PP)} \Hr{\alpha}(\QQ), &0\leq\alpha<1, \\
    \sup_{\QQ\in\ball(\PP)} \Hr{\alpha}(\QQ), &1<\alpha\leq\infty.
    \end{array}\right.
\]
\end{definition}
At the end of this paper, we point out that $\Hrs{\alpha}{\epsilon}(\PP)$
will actually vary, depending on which notion of $\epsilon$-closeness is used,
leading to a maximum difference of $\frac{\alpha}{\alpha-1}\log(1-\epsilon)$.

For a probability distribution $\PP$ on, e.g., $\mathcal{Z}=\{0,1\}^{\mN}$, we
define $\PP^n$ as the probability distribution
corresponding to the restriction of the ``infinite volume'' distribution
$\PP$ to the finite volume $\{0,\ldots,n-1\}$.
\begin{definition}[Entropy rate of a classical source]\label{def:classicrate}
For a stationary source given by its probability measure $\PP$, we
define
\begin{eqnarray*}
\h(\PP) &=&
    \lim_{n\rightarrow\infty} \frac{1}{n} \H(\PP^n),
\\
\hrs{\alpha}{\epsilon}(\PP) &=&
    \lim_{n\rightarrow\infty} \frac{1}{n}\Hrs{\alpha}{\epsilon}(\PP^n).
\end{eqnarray*}
\end{definition}
We will actually prove that $\hrs{\alpha}{\epsilon}(\PP)=\h(\PP)$ as $\epsilon\rightarrow0$.

We use the standard notion of typical sequences and typical sets, which are defined for any information
source (not necessarily i.i.d.). See, for instance, \cite{CV06} or \cite{NC00}.
\begin{definition}[Typical sequences, typical set]
A sequence $z^n\in\bit^n$, $n\in\mN$, is called $\epsilon$-typical if
\[  \exp{-n(h(\PP)+\epsilon)} \leq \PP(z^n) \leq \exp{-n(h(\PP)-\epsilon)}. \]
The typical set $\typ$ is the set of all $\epsilon$-typical sequences from $\bit^n$.
\end{definition}

In this paper we need the following consequence of the AEP, where we refer to \cite[Section~16.8]{CV06} for the AEP
for ergodic sources (known as the Shannon-McMillan-Breiman theorem).
\begin{theorem}[Classical AEP bounds] \label{thm:classicalaep}
Let $\PP$ be a stationary, ergodic probability distribution on $\mathcal{Z}=\{0,1\}^{\mN}$.
Let $\epsilon>0$. Then, for sufficiently large $n$,
\[\PP(\typ)\geq1-\epsilon, \]
and
\[ \card{\typ}\leq \exp{n(h(\PP)+\epsilon)}. \]
\end{theorem}

\begin{definition}[Quantum R\'enyi entropy]
The R\'enyi entropy of order $\alpha\in[0,\infty]$ of a density matrix $\rho$ is
\[
\Sr{\alpha}(\rho) = \frac{1}{1-\alpha} \log \tr(\rho^\alpha)
\]
for $0<\alpha<\infty$, $\alpha\neq1$, and
$\Sr{\alpha}(\rho) =\lim_{\beta\rightarrow\alpha} \Sr{\beta}(\rho)$ otherwise.
\end{definition}
Hence, $\Sr{0}(\rho) = \log\rank(\rho)$, $\Sr{1}(\rho)=\S(\rho)=-\tr(\rho \log \rho)$
(von Neumann entropy) and $\Sr{\infty}(\rho) = - \log \lambda_{\mathrm max}(\rho)$.

Analogous to the classical case, smooth R\'enyi entropy is defined in the quantum
case (see \cite{RK05}). We use either the set of density matrices which are $\epsilon$-close to $\rho$,
$\ball(\rho)=\{\sigma: \delta(\rho,\sigma)\leq\epsilon\}$ or the set of
non-normalized density matrices which are $\epsilon$-close to $\rho$,
$\ball(\rho)=\{\sigma: \tr(\sigma)\geq1-\epsilon,
0\leq\sigma\leq\rho\}$. The first notion of $\epsilon$-closeness, based on the
trace distance $\delta(\rho,\sigma) = \frac{1}{2}\tr(|\rho-\sigma|)$, was used in \cite{RK05}.
The second notion is introduced here, and will actually be used in the next section.
\begin{definition}[Quantum smooth R\'enyi entropy, \cite{RK05}]\label{def:quantumsmooth}
The $\epsilon$-smooth R\'enyi entropy of order $\alpha\in[0,1)\cup(1,\infty]$ of a density
matrix $\rho$ is
\[
\Srs{\alpha}{\epsilon}(\rho) = \left\{\begin{array}{ll}
    \inf_{\sigma\in\ball(\rho)} \Sr{\alpha}(\sigma), &0\leq\alpha<1, \\
    \sup_{\sigma\in\ball(\rho)} \Sr{\alpha}(\sigma), &1<\alpha\leq\infty.
    \end{array}\right.
\]
\end{definition}

\begin{definition}[Entropy rates of a quantum source]\label{def:quantumrate}
For a stationary quantum source $\rho$, given by
its local densities $\n{\rho}=\rho_{0,\ldots,n-1}$, for $n\in\mN$,
we define:
\begin{eqnarray*}
\s(\rho) &=&
    \lim_{n\rightarrow\infty} \frac{1}{n} \S(\n{\rho}),
\\
\srs{\alpha}{\epsilon}(\rho) &=&
    \lim_{n\rightarrow\infty} \frac{1}{n}\Srs{\alpha}{\epsilon}(\n{\rho}).
\end{eqnarray*}
\end{definition}

We use the following notion of typical states and typical subspaces, as can be found
in \cite{BS03} (also see \cite{NC00}).
\begin{definition}[Typical state, typical subspace]
A pure state $|\n{e_i}\rangle$, where $\n{e_i}$ is an eigenvector of $\n{\rho}$
is called $\epsilon$-typical if the corresponding eigenvalue $\n{\lambda_i}$ satisfies
\[  \exp{-n(s(\rho)+\epsilon)} \leq \n{\lambda_i} \leq \exp{-n(s(\rho)-\epsilon)}. \]
The typical subspace $\typ$ is the subspace spanned by all $\epsilon$-typical states.
\end{definition}

We will need the following consequences of the quantum AEP for ergodic sources, which
has been studied in \cite{BS03} (see \cite{NC00} for the quantum AEP for i.i.d.\ sources).
\begin{theorem}[Quantum AEP bounds] \label{thm:quantumaep}
Let $\rho$ be a stationary, ergodic quantum source with
local densities $\n{\rho}$. Let $\epsilon>0$. Then, for sufficiently large $n$,
\[ \tr(\n{\rho}P_{\mathcal{T}_\epsilon^{(n)}}) \geq 1-\epsilon, \]
where $P_{\typ}$ is the projector onto the subspace $\typ$.
Furthermore,
\[ \tr(P_{\typ}) \leq \exp{n(\s(\rho)+\epsilon)}.\]
\end{theorem}
Clearly, the quantum AEP for ergodic sources implies the classical AEP for ergodic sources.

The following theorem by Renner and Wolf states that smooth R\'enyi
entropy approaches Shannon entropy in the case of a classical
i.i.d.\ source.
\begin{theorem}[\protect{\cite[Lemma~I.2]{RW05}}]\label{thm:RW05}
Let $Z^n$ denote an $n$-tuple of i.i.d.\ random variables with probability distribution $\PP_Z$. Then,
\[ \lim_{\epsilon\rightarrow0} \lim_{n\rightarrow\infty} \frac{1}{n}\Hrs{\alpha}{\epsilon}(Z^n) = \H(Z), \]
for any $\alpha\in[0,\infty]$.
\end{theorem}

The analogous theorem by Renner and K\"onig for a quantum i.i.d.\ source is as follows.
\begin{theorem}[\protect{\cite[Lemma~3]{RK05}}]\label{thm:RK05}
Let $\rho$ be a density matrix. Then,
\[ \lim_{\epsilon\rightarrow0} \lim_{n\rightarrow\infty} \frac{1}{n}\Srs{\alpha}{\epsilon}(\rho^{\otimes n})  = \S(\rho), \]
for any $\alpha\in[0,\infty]$.
\end{theorem}

\section{Main Result}
We extend the results by Renner and Wolf (Theorem~\ref{thm:RW05} above) and by Renner and
K\"onig (Theorem~\ref{thm:RK05} above) to the case of ergodic sources. Throughout this section,
we use the notion of $\epsilon$-closeness based on non-normalized probability distributions
and density matrices, so $\ball(\PP)=\{\QQ: \sum_{z\in\mathcal{Z}}\QQ(z)\geq1-\epsilon,
\forall_{z\in\mathcal{Z}} 0\leq \QQ(z) \leq \PP(z)\}$ and $\ball(\rho)=\{\sigma: \tr(\sigma)\geq1-\epsilon,
0\leq\sigma\leq\rho\}$, respectively. In the next section, we will argue that the results
are independent on which notion of $\epsilon$-closeness is used.

\subsection{Classical Case}
We start with our main result for the classical case. The known result
for an i.i.d.\ source is by Renner and Wolf, Theorem~\ref{thm:RW05} above.
We will extend this to a stationary, ergodic source in Theorem~\ref{thm:classical} below.

\begin{lemma}\label{lemma:clasmin}
Let $\PP$ be a stationary, ergodic information source given by its probability measure
and let $0<\epsilon < 1/2$. Then we have,
\[\h(\PP)-\epsilon \leq \hrs{\infty}{\epsilon}(\PP) \leq  \h(\PP)+2\epsilon.\]
\end{lemma}

\proof Let $0<\epsilon < 1/2$. To prove the lower bound, we show
that, for sufficiently large $n$, $\Hrs{\infty}{\epsilon}(\PP^n) \geq n(\h(\PP)-\epsilon)$.
Define non-normalized probability distribution $\QQ$ for all $z^n\in\bit^n$ by
\begin{equation}\label{eq:truncatedP}
\QQ(z^n)=\left\{
\begin{array}{cl}
\PP(z^n), &\textrm{if } z^n\in\typ\\
0,        &\textrm{if } z^n\notin\typ.
\end{array}\right.
\end{equation}
Clearly, $0\leq \QQ(z^n)\leq\PP(z^n)$ and,
by the AEP, $\QQ(\typ)=\PP(\typ)\geq 1-\epsilon$
for sufficiently large $n$. So,
$\QQ \in\ball(\PP^n)$.
Furthermore, for $z^n\in\typ$, we have that
$-\log \PP(z^n) \geq n(\h(\PP)-\epsilon)$,
and hence that for any $z^n$ that
$-\log \QQ(z^n) \geq n(\h(\PP)-\epsilon)$.
This implies that $\Hr{\infty}(\QQ)=-\log \max_{z^n} \QQ(z^n)\geq n(\h(\PP)-\epsilon)$
and the lower bound follows.

Next, to prove the upper bound, we show that,
for sufficiently large $n$, one has that for all
$\QQ\in\ball(\PP^n)$,
\[
\Hr{\infty}(\QQ) =- \log \max_{z^n} \QQ(z^n)\leq n(h(\PP)+2\epsilon).
\]
This follows from $\max_{z^n\in\typ} \QQ(z^n)\geq \exp{-n(h(\PP)+2\epsilon)}$,
which in turn follows from $\sum_{z^n\in\typ} \QQ(z^n)\geq \card{\typ} \exp{-n(h(\PP)+2\epsilon)}$.
From the AEP we get $\card{\typ}\leq \exp{n(h(\PP)+\epsilon)}$, hence it suffices
to prove that, for sufficiently large $n$,
\begin{equation}\label{eq:typsumQ}
\sum_{z^n\in\typ} \QQ(z^n)\geq \exp{-n\epsilon}.
\end{equation}
As $\sum_{z^n} \QQ(z^n)\geq 1-\epsilon$, for $\QQ\in\ball(\PP^n)$,
and also $\sum_{z^n\notin\typ} \QQ(z^n)\leq \epsilon$ (because $\QQ(z^n)\leq \PP(z^n)$
and $P(\typ)\geq 1-\epsilon$ from the AEP), we only need to observe that
\[
1-2\epsilon > e^{-n\epsilon}
\]
holds for sufficiently large $n$, using that $\epsilon<1/2$.
\endproof

We now state an analogous lemma for the {\em max}-entropy.
\begin{lemma}\label{lemma:clasmax}
Let $\PP$ be a stationary, ergodic information source given by its probability measure
and let $0<\epsilon < 1/2$. Then we have,
\[\h(\PP)-2\epsilon \leq \hrs{0}{\epsilon}(\PP) \leq  \h(\PP)+\epsilon.\]
\end{lemma}
\proof Let $0<\epsilon < 1/2$. To prove the upper bound, we show
that, for sufficiently large $n$, $\Hrs{0}{\epsilon}(\PP^n) \leq n(\h(\PP)+\epsilon)$.
We do so by showing that $\Hr{0}(\QQ)=\log \card{\{z^n:  \QQ(z^n)>0\}} \leq n(\h(\PP)+\epsilon)$
for the non-normalized probability distribution $\QQ$, defined by (\ref{eq:truncatedP}) in the proof
of Lemma~\ref{lemma:clasmin}. As
\[ \card{\{z^n:  \QQ(z^n)>0\}} = \card{\{z^n\in\typ :  \QQ(z^n)>0\}} \leq \card{\typ}, \]
the result follows directly from the AEP.

Next, to prove the lower bound, we show that,
for sufficiently large $n$, one has that for all
$\QQ\in\ball(\PP^n)$,
\[
\Hr{0}(\QQ) =\log \card{\{z^n:  \QQ(z^n)>0\}}\geq n(h(\PP)-2\epsilon).
\]
This is implied by $\card{\{z^n\in\typ:  \QQ(z^n)>0\}}\geq \exp{n(h(\PP)-2\epsilon)}$,
which is in turn implied by
\[ \sum_{z^n\in\typ} \QQ(z^n)\geq \max_{z^n\in\typ, \QQ(z^n)>0} \QQ(z^n) \exp{n(h(\PP)-2\epsilon)}.\]
Using inequality (\ref{eq:typsumQ}) from the proof of Lemma~\ref{lemma:clasmin}, it suffices
to show that
\[ \max_{z^n\in\typ, \QQ(z^n)>0} \QQ(z^n) \leq \exp{-n\epsilon}\exp{-n(h(\PP)-2\epsilon)} = \exp{-n(h(\PP)-\epsilon)}.\]
This is a direct consequence of the definition of $\epsilon$-typical sequences, as
$\QQ(z^n)\leq\PP(z^n)$ for $\QQ\in\ball(\PP^n)$, using that $\QQ(z^n)>0$ holds for at least one $z^n\in\typ$
on account of inequality (\ref{eq:typsumQ}).
\endproof

\begin{theorem}\label{thm:classical}
For $\alpha\in[0,\infty]$, the $\epsilon$-smooth entropy of a stationary, ergodic
information source $\PP$ given by its probability measure on ${\cal{Z}}=\{0,1\}^\mathbb{N}$ is
close to the mean Shannon entropy:
\[ \lim_{\epsilon\rightarrow0} \hrs{\alpha}{\epsilon}(\PP) = \h(\PP) .\]
\end{theorem}
\proof
For $\alpha<1$, the monotonicity of smooth R\'enyi entropy (see, e.g., \cite[Lemma~1]{RW05}) yields
$\Hrs{\alpha}{\epsilon}(\PP^n)\leq \Hrs{0}{\epsilon}(\PP^n)$, and hence
$\hrs{\alpha}{\epsilon}(\PP)\leq\h(\PP)+\epsilon$ by Lemma~\ref{lemma:clasmax}.

To get a lower bound for $\hrs{\alpha}{\epsilon}(\PP)$, we note that
\[ \Hrs{\alpha}{\epsilon}(\PP^n)\geq \Hrs{0}{2\epsilon}(\PP^n)-\frac{\log(1/\epsilon)}{1-\alpha}, \]
using \cite[Lemma~2]{RW05}.
So, $\hrs{\alpha}{\epsilon}(\PP)\geq \hrs{0}{2\epsilon}(\PP)$ as the constant
term on the right-hand side vanishes for $n\rightarrow\infty$.
Using Lemma~\ref{lemma:clasmax}, we thus get $\hrs{\alpha}{\epsilon}(\PP)\geq \h(\PP)-4\epsilon$.

This proves that $\lim_{\epsilon\rightarrow0} \hrs{\alpha}{\epsilon}(\PP) = \h(\PP)$.
The proof for $\alpha>1$ is completely symmetrical, hence omitted.
\endproof
Note that the term $2\epsilon$ in the upper and lower bounds of Lemmas~\ref{lemma:clasmin}
and~\ref{lemma:clasmax}, respectively, can be improved to $(1+\delta)\epsilon$ for any constant $\delta>0$. Similarly, the term
$4\epsilon$ in the proof of Theorem~\ref{thm:classical} for the lower bound for $\hrs{\alpha}{\epsilon}$ can be improved to
$(1+\delta)\epsilon$ for any constant $\delta>0$.

\subsection{Quantum Case}
Although it is possible to prove the quantum case directly, along the same lines as in the classical case,
we treat the quantum case indirectly, by reducing it to the classical case. This leads to a more compact proof.
To this end, we will first
prove Lemma~\ref{lemma:newballquantumclassical} below, which captures the correspondence between $\ball(\n{\rho})$ and $\ball(\n{\lambda})$. We only consider the case of $\epsilon$-closeness for non-normalized density matrices and probability distributions
(but the lemma also holds for the case of $\epsilon$-closeness based on trace distance).

To prove our lemma, we need Weyl's monotonicity principle which we recall first.
\begin{theorem}[Weyl monotonicity] \label{thm:weyl}
If $A$, $B$ are $m$ by $m$ Hermitian matrices and $B$ is positive, then
$\lambda_i(A) \leq \lambda_i(A+B)$ for all $i=1,\ldots,m$, where
$\lambda_i(M)$ is the $i$-th eigenvalue of $M$ (ordered from largest to smallest).
\end{theorem}
\begin{lemma} \label{lemma:newballquantumclassical}
Let $\rho$ be a density matrix with eigenvalues $\lambda_1\geq\lambda_2\geq\ldots\geq\lambda_m$.
\begin{enumerate}
\item
For any density matrix $\sigma$ with eigenvalues $\mu_1\geq\mu_2\geq\ldots\geq\mu_m$,
\[
\sigma \in \ball(\rho) \Rightarrow \mu \in \ball(\lambda).
\]
\item
Given real numbers $\mu_1,\ldots,\mu_m$ such that $\mu\in \ball(\lambda)$, there
exists a matrix $\sigma$ with eigenvalues $\mu_1,\ldots,\mu_m$ such
that $\sigma \in \ball(\rho)$.
\end{enumerate}
\end{lemma}
\proof We prove the result for
\begin{eqnarray*}
\ball(\lambda) &=&
    \{\mu:  \sum_i \mu_i\geq1-\epsilon, \forall_i 0\leq \mu_i \leq \lambda_i\}, \\
\ball(\rho) &=&
 \{\sigma: \tr(\sigma)\geq1-\epsilon, 0\leq\sigma\leq\rho\}.
\end{eqnarray*}
For the first part, let $\sigma$ be a (possibly non-normalized) density matrix with
eigenvalues $\mu_1\geq\mu_2\geq\ldots\geq\mu_m$ and suppose $\sigma \in \ball(\rho)$.
Since $\sigma$ is positive we have $\mu_i \geq 0$ for all $i$.
And since $\sigma\leq\rho$, we have that $\rho-\sigma$ is positive as well, so
$\lambda_i\geq\mu_i$ for all $i$ (using Weyl's monotonicity principle, Theorem~\ref{thm:weyl} above).
Finally, note that $\tr(\sigma)\geq1-\epsilon$ is equivalent to $\sum_i \mu_i \geq 1-\epsilon$,
so we conclude that $\mu \in \ball(\lambda)$.

For the second part, let $\mu\in \ball(\lambda)$ be given.
We write the Hermitian matrix $\rho$ in diagonal form,
\[ \rho=\sum_i \lambda_i|v_i\rangle \langle v_i|.\]
for eigenvectors $v_i$ ($i=1,\ldots,m$), and
we show that the Hermitian matrix $\sigma$, defined by
\[ \sigma=\sum_i \mu_i|v_i\rangle \langle v_i|,\]
is in $\ball(\rho)$.

Since $\mu\in\ball(\lambda)$, we have that $0\leq \mu_i\leq\lambda_i$, and because
$\rho$ and $\sigma$ commute (eigenvalues of $\rho-\sigma$ are $\lambda_i-\mu_i$), we have $0\leq\sigma\leq\rho$.
Clearly, $\sum_i \mu_i \geq 1-\epsilon$ so $\tr(\sigma)\geq1-\epsilon$ as well, and
therefore $\sigma \in \ball(\rho)$.
\endproof

We now proceed to prove the main result for the quantum case.
\begin{theorem}\label{thm:quantum}
For $\alpha\in[0,\infty]$, the $\epsilon$-smooth entropy of a stationary, ergodic
quantum source $\rho$ given by its local densities $\n{\rho}$, for $n\in\mN$, is
close to the mean von Neumann entropy:
\[ \lim_{\epsilon\rightarrow0} \srs{\alpha}{\epsilon}(\rho) = \s(\rho) .\]
\end{theorem}
\proof We will apply Theorem~\ref{thm:classical} as follows.

First note that for the local densities $\n{\rho}$ for a quantum information source $\rho$,
we have that $\S(\n{\rho})=\H(\n{\lambda})$, where $\n{\lambda}$ denotes the probability distribution corresponding
to the eigenvalues of $\n{\rho}$. Consequently, $\s(\rho)=\h(\lambda)$ as well, where
$\lambda$ denotes the probability distribution corresponding to the eigenvalues of $\rho$.

Next, we recall the definitions of smooth R\'enyi entropy in the classical and quantum case, resp.:
\begin{eqnarray*}
\Hrs{\alpha}{\epsilon}(\PP) &=&
    \left\{\begin{array}{ll}
    \inf_{\QQ\in\ball(\PP)} \Hr{\alpha}(\QQ), &0\leq\alpha<1, \\
    \sup_{\QQ\in\ball(\PP)} \Hr{\alpha}(\QQ), &1<\alpha\leq\infty.
    \end{array}\right.
\\
\Srs{\alpha}{\epsilon}(\rho) &=&
    \left\{\begin{array}{ll}
    \inf_{\sigma\in\ball(\rho)} \Sr{\alpha}(\sigma), &0\leq\alpha<1, \\
    \sup_{\sigma\in\ball(\rho)} \Sr{\alpha}(\sigma), &1<\alpha\leq\infty.
    \end{array}\right.
\end{eqnarray*}
We only consider the case $\alpha<1$, as the other case follows by symmetry.
We have that
\begin{eqnarray*}
 \Srs{\alpha}{\epsilon}(\n{\rho}) &=&   \inf_{\sigma\in\ball(\n{\rho})} \Sr{\alpha}(\sigma) \\
  &=&  \inf_{\mu\in\ball(\n{\lambda})} \Hr{\alpha}(\mu) \\
  &=& \Hrs{\alpha}{\epsilon}(\n{\lambda}),
\end{eqnarray*}
using that Lemma~\ref{lemma:newballquantumclassical} implies that the infimum over
$\ball(\n{\rho})$ is equal to the infimum over $\ball(\n{\lambda})$.

As a consequence, we have that $\srs{\alpha}{\epsilon}(\rho)=\hrs{\alpha}{\epsilon}(\lambda)$
and the result follows from Theorem~\ref{thm:classical}. Here, we use the fact that quantum AEP
implies classical AEP.
\endproof
We note that the actual convergence rate (as a function of $\epsilon$) is the same as in the classical case,
which follows by considering the analogons of Lemmas~\ref{lemma:clasmin} and~\ref{lemma:clasmax}.

\section{Notions of $\epsilon$-Closeness}
As mentioned in the introduction, two notions of $\epsilon$-closeness were originally introduced
by Renner and Wolf \cite{RW04,RW05}, which can both be used in the definition of classical
smooth R\'enyi entropy. For the quantum case, the
paper by Renner and K\"onig \cite{RK05} only considers the notion of $\epsilon$-closeness based on the
trace distance. As the natural quantum analogon of the notion of $\epsilon$-closeness based
on non-normalized probability distributions, we have used the set of non-normalized
density matrices which are $\epsilon$-close to a given density matrix $\rho$:
\[ \ball(\rho)=\{\sigma: \tr(\sigma)\geq1-\epsilon, 0\leq\sigma\leq\rho\}. \]

The entropy rates (Definitions~\ref{def:classicrate} and~\ref{def:quantumrate}), and
consequently the results for these entropy rates (Theorems~\ref{thm:RW05}, \ref{thm:RK05},
\ref{thm:classical}, and~\ref{thm:quantum}) do not depend on which of these notions of
$\epsilon$-closeness is used.

Furthermore, if the corresponding notions of $\epsilon$-closeness are used,
the quantum case and the classical case are in general connected as follows:
\[  \Srs{\alpha}{\epsilon}(\rho) =
 \inf_{\sigma\in\ball(\rho)} \Sr{\alpha}(\sigma) =
 \inf_{\mu\in\ball(\lambda)} \Hr{\alpha}(\mu) = \Hrs{\alpha}{\epsilon}(\lambda),\]
where $\lambda$ denotes the probability distribution corresponding to the eigenvalues of $\rho$.

We note, however, that the smooth R\'enyi entropy $\Hrs{\alpha}{\epsilon}$ may depend
on which notion of $\epsilon$-closeness is used, contrary to what was stated before
(see, e.g., Section~3.3 of the full version of \cite{RW04}). In general, one can show
that
\[ 0\leq \inf_{\delta(\PP,\QQ)\leq\epsilon} \Hr{\alpha}(\QQ) - \inf_{\stackrel{\sum_{z}\QQ(z)\geq1-\epsilon}{
\forall_{z} 0\leq \QQ(z) \leq \PP(z)}} \Hr{\alpha}(\QQ) \leq \frac{\alpha}{\alpha-1} \log (1-\epsilon), \]
for $0\leq\alpha<1$, and that
\[ \frac{\alpha}{\alpha-1} \log (1-\epsilon)\leq \sup_{\delta(\PP,\QQ)\leq\epsilon} \Hr{\alpha}(\QQ) - \sup_{\stackrel{\sum_{z}\QQ(z)\geq1-\epsilon}{
\forall_{z} 0\leq \QQ(z) \leq \PP(z)}} \Hr{\alpha}(\QQ) \leq 0, \]
for $1<\alpha\leq\infty$.
So, only for $\alpha=0$ either notion of $\epsilon$-closeness yields the same value for the
smooth R\'enyi entropy $\Hrs{\alpha}{\epsilon}$. But for all other values of $\alpha$, the difference
may be as large as $\frac{\alpha}{\alpha-1} \log (1-\epsilon)$. The maximum difference is attained
for the uniform distribution $\PP(z)=1/m$ on a finite range $\cal Z$ of size $m$, assuming that
$\epsilon$ is sufficiently small (i.e., $\epsilon<1/m$).


\section*{Acknowledgment}
Boris \v{S}kori\'c is gratefully acknowledged for discussions in the early stage of this work.

\end{document}